\documentclass[twocolumn,aps,prb]{revtex4}
\usepackage{amsmath}

\usepackage{amsfonts,amsxtra,graphicx,float} 

\begin{document}

\title{A WKB--like approach to Unruh Radiation} 

\author{Andrea~de~Gill}
\email{aadegill@csufresno.edu} \affiliation{Physics Department,
California State University Fresno, Fresno, California 93740-8031}

\author{Douglas~Singleton}
\email{dougs@csufresno.edu}
\affiliation{Physics Department, CSU Fresno, Fresno, CA 93740-8031 \\
and \\
Institute of Gravitation and Cosmology, Peoples' Friendship University of Russia, Moscow 117198, Russia}

\author{Valeria~Akhmedova}
\email{lera@itep.ru} \affiliation{ITEP, B. Cheremushkinskaya, 25,
Moscow, Russia 117218}

\author{Terry~Pilling}
\email{Terry.Pilling@ndsu.edu} \affiliation{Department of Physics,
North Dakota State University, Fargo, North Dakota 58105-5566}

\date{\today} 

\begin{abstract}

Unruh radiation is the thermal flux seen by an accelerated observer moving through Minkowski spacetime. In this 
article we study Unruh radiation as tunneling through a barrier. We use a WKB--like method to obtain the tunneling rate 
and the temperature of the Unruh radiation. This derivation brings together many topics into a single problem -- classical mechanics, relativity, relativistic field theory,
quantum mechanics, thermodynamics and mathematical physics. Moreover, this gravitational WKB
method helps to highlight the following subtle points: (i) the tunneling rate strictly should be written as the closed path integral of the canonical momentum; (ii) for the case of the gravitational WKB problem, there is a time--like contribution to the tunneling rate arising from an imaginary change of the time coordinate upon crossing the horizon. This temporal contribution to the tunneling rate has no analog in the ordinary quantum mechanical WKB calculation.

\end{abstract} 

\maketitle
%%%%%%%%%%%%%%%%%%%%%%%%%%%%%%%%%%%%%%%%%%%%%%%%%%% 

\section{Introduction} 

The radiation that arises from placing a quantum field in a background metric 
with a horizon is a well known phenomenon at the boundary between field theory 
and general relativity. The first example of this effect was Hawking radiation 
\cite{hawking}, where a Schwarzschild black hole radiates with a thermal spectrum 
at the expense of the black hole's mass. Another example is Hawking--Gibbons radiation 
\cite{gibbons}, i.e., the thermal radiation seen by an observer in de Sitter spacetime. 
In this paper we focus on Unruh radiation \cite{unruh} -- the radiation seen by an 
observer moving with a constant acceleration through vacuum. The original methods 
used to calculate these effects used quantum field theory at a level which is 
beyond most undergraduates or beginning graduate students. In reference \onlinecite{hawking}, 
Hawking gave a heuristic picture for the radiation in terms of ``tunneling" of virtual 
particles across the horizon. After a span of twenty five years, mathematical details
were added to this picture \cite{kraus, padman, parikh1, parikh2}. In these works, the 
action for a particle which crosses the horizon of some spacetime (e.g., the Schwarzschild 
spacetime for the case of Hawking radiation) was calculated and found to have an imaginary 
part coming from a contour integration. The exponential of this imaginary piece was 
compared to a Boltzmann distribution, which allowed one to determine the
temperature of the radiation. The simplicity of this gravitational WKB method makes it 
easy to calculate Hawking like radiation for the case of other metrics (e.g. Reissner--Nordstrom 
\cite{parikh1}, de Sitter\cite{sekiwa, volovik, medved, temporal}, Kerr and Kerr--Newmann 
\cite{kerr, kerr2}, Unruh \cite{unruhPLB}). Additionally, one could easily incorporate 
tunneling particles with different spins \cite{mann} and one could 
(in a simplified way) begin to take into account back reaction effects on the metric 
\cite{parikh1, parikh2, vagenas}.
   
In reference \onlinecite{donoghue}, Unruh radiation is derived using purely quantum mechanical arguments. However, the reader needs to know the quantized radiation field, and the mathematical steps in the derivation are more involved as compared to the approach presented here. In comparison with reference \onlinecite{donoghue}, the gravitational WKB--like method is mathematically simple while at the same time it provides a clear physical picture for the origins of the radiation. In this article, this WKB--like method is presented in a pedagogical manner for the case of the Rindler spacetime (the metric seen by an observer who undergoes constant proper acceleration) 
and Unruh radiation. The reason for choosing Rindler spacetime is that it is 
the simplest spacetime in which this effect occurs. Furthermore, because
of the strong equivalence principle (i.e., locally, a constant acceleration and a gravitational field are
observationally equivalent), the Unruh radiation from Rindler spacetime is the prototype of this
type of effect. Also, of all these effects -- Hawking radiation, Hawking--Gibbons radiation --
Unruh radiation has the best prospects for being observed experimentally \cite{jackson, bell, akhmedov1, retz}.

This derivation of Unruh radiation draws together many different areas of study: (i) classical mechanics via the Hamilton--Jacobi equations; (ii) relativity via the use of the 
Rindler metric; (iii) relativistic field theory through the Klein--Gordon equation in curved backgrounds; (iv) quantum mechanics via the use of the WKB--like
method applied to gravitational backgrounds; (v) thermodynamics via the use of the Boltzmann
distribution to extract the temperature of the radiation; (vi) mathematical methods in physics via the
use of contour integrations to evaluate the imaginary part of the action of the particle that crosses
the horizon. Thus this single problem serves to show students how the different
areas of physics are interconnected.

Also, through this example we will highlight some subtle features of the Rindler metric and the WKB
method which are usually overlooked. In particular, we show that the gravitational WKB amplitude has a contribution
coming from a change of the time coordinate from crossing the horizon \cite{temporal}. This temporal contribution is
never encountered in ordinary quantum mechanics, where time acts as a parameter rather than
a coordinate.        

\section{Rindler spacetimes}

In this section we introduce and discuss some relevant features of Rindler spacetime -- the spacetime
seen by an observer  moving with constant proper acceleration through Minkowski spacetime. 
The Rindler metric can be obtained by starting with the Minkowski metric, i.e., $ds^2 = -dt^2 + dx^2 + dy^2 + dz^2$, 
where we have set $c=1$, and transforming to the coordinates of the accelerating observer. We take the acceleration 
to be along the $x$--direction, thus we only need to consider a 1+1 dimensional Minkowski spacetime
\begin{equation}
\label{flat}
ds^2 = -dt^2 + dx^2~.
\end{equation}
Using the Lorentz transformations (LT) of special relativity, the 
worldlines of an accelerated observer moving along the $x$--axis in empty spacetime can be related 
to Minkowski coordinates $t$, $x$ according to the following transformations
\begin{equation}
\label{rindcoor}
\begin{split}
t &= (a^{-1} + x_R)\sinh(at_R) \\
x &= (a^{-1} + x_R)\cosh(at_R) ~,
\end{split}
\end{equation}
where $a$ is the constant, proper acceleration of the Rindler observer measured in his instantaneous rest frame. 
One can show that the acceleration associated with the trajectory of
\eqref{rindcoor} is constant since $a_\mu a^\mu = (d^2 x_\mu / dt_R ^2)^2 = a^2$ with 
$x_R =0$. The trajectory of \eqref{rindcoor} can be obtained using the definitions of 
four--velocity and four--acceleration of the accelerated observer in his instantaneous inertial 
rest frame \cite{MTW}.  Another derivation of \eqref{rindcoor} uses a LT to relate the proper acceleration of the 
non--inertial observer to the acceleration of the inertial observer \cite{rindlerSR}. The text by Taylor and Wheeler \cite{taylor} also provides a discussion of the Rindler observer.

The coordinates $x_R$ and $t_R$, when parametrized and plotted in a spacetime diagram whose axes 
are the Minkowski coordinates $x$ and $t$, result in the familiar hyperbolic trajectories (i.e., $x^2 - t^2 = a^{-2}$) 
that represent the worldlines of the Rindler observer.

Differentiating each coordinate in \eqref{rindcoor} and substituting the result 
into \eqref{flat} yields the standard Rindler metric
\begin{equation}
\label{rindler}
ds^2 = -(1 + a x_R)^2 dt_R^2 + dx_R^2 ~.
\end{equation}
When $x_{R}= -\frac{1}{a}$, the determinant of the metric given by \eqref{rindler}, $det(g_{ab})\equiv  g=-(1 + a x_R)^2$, 
vanishes. This indicates the presence of a coordinate singularity at $x_{R}= -\frac{1}{a}$, which can not be a real
singularity since \eqref{rindler} is the result of a global coordinate transformation from Minkowski spacetime. The horizon of the Rindler
spacetime is given by $x_{R}= -\frac{1}{a}$. 

\begin{figure} [H]
  \centering\includegraphics[trim = 50mm 80mm 50mm 80mm, clip, width=8.5cm]{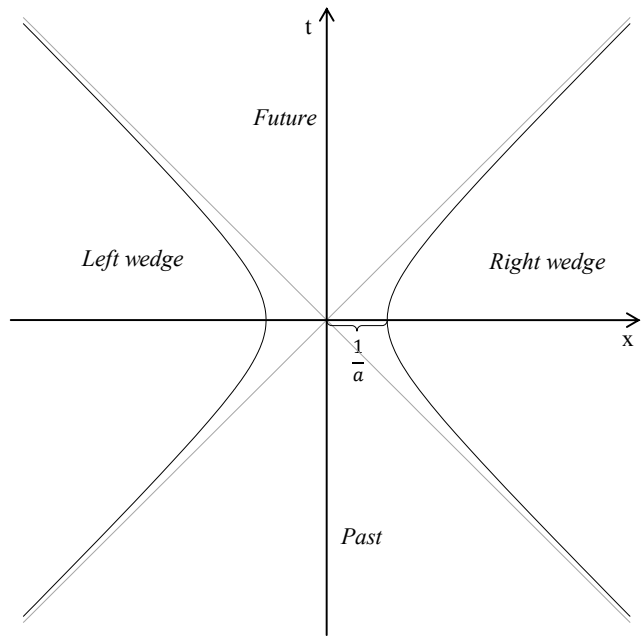}
\noindent\caption{Trajectory of the Rindler observer as seen by the observer at rest.}
\label{fig:traj}
\end{figure}

In the spacetime diagram shown above, the horizon for this metric is represented by the null asymptotes, 
$x = \pm t$,  that the hyperbola given by \eqref{rindcoor} approaches as $x$ and $t$ tend 
to infinity \cite{BD}. Note that this horizon is a particle horizon, since the Rindler observer is not influenced by the whole spacetime, and the horizon's location is observer dependent \cite{visser}.

One can also see that the transformations \eqref{rindcoor} that lead to the Rindler metric
in \eqref{rindler} only cover a quarter of the full Minkowski spacetime, given by $x-t>0$ and $x+t>0$. This portion of Minkowski is usually labeled \textit{Right wedge}. To recover the \textit{Left wedge}, one can modify the second equation of \eqref{rindcoor} with a minus sign in front of the transformation of the $x$ coordinate, thus recovering the trajectory of an observer moving with a negative acceleration. In fact, 
we will show below that the coordinates $x_R$ and $t_R$ double cover the region in front of the horizon,  $x_{R}= -\frac{1}{a}$. In this sense, the metric in \eqref{rindler} is similar to the Schwarzschild metric written in isotropic coordinates. For further details, see reference \onlinecite{visser}.

There is an alternative form of the Rindler metric that can be
obtained from \eqref{rindler} by the following transformation:
\begin{equation}
\label{coordtrans}
(1+ a \, x_{R})= \sqrt{| 1+ 2\,a\,x_{R'} |} ~.
\end{equation}
Using the coordinate transformation given by \eqref{coordtrans} in \eqref{rindler}, we get the following 
Schwarzschild--like form of the Rindler metric
\begin{equation}
\label{rindler2}
ds^2 = -(1 + 2\,a\,x_{R'})dt_{R'}^2 + (1 + 2\,a\,x_{R'})^{-1} dx_{R'}^2 ~.
\end{equation}
If one makes the substitution $a \rightarrow GM/x_{R'}^2$ one can see the similarity to
the usual Schwarzschild metric. The horizon is now at
$x_{R'} = -1/2a$ and the time coordinate, $t_{R'}$, does change sign
as one crosses $x_{R'} = -1/2a$. In addition, from \eqref{coordtrans} one can see explicitly
that as $x_{R'}$ ranges from $+\infty$ to $-\infty$ the standard Rindler coordinate will
go from $+\infty$ down to $x_R = -1/a$ and then back out to $+\infty$.

The Schwarzschild--like form of the Rindler metric given by \eqref{rindler2} can also be obtained 
directly from the $2$--dimensional Minkowski metric \eqref{flat} via the transformations
\begin{equation}
\begin{split}
\label{coordtrans1}
t &= \frac{\sqrt{1+2ax_{R'}}}{a} \sinh (at_{R'}) \\ 
x &= \frac{\sqrt{1+2ax_{R'}}}{a} \cosh (at_{R'})  
\end{split}
\end{equation}
\noindent for $x_{R'} \geq -\frac{1}{2a}$, and
\begin{equation}
\begin{split}
\label{coordtrans2}
t &= \frac{\sqrt{| 1+2ax_{R'} |}}{a} \cosh (at_{R'}) \\
x &= \frac{\sqrt{| 1+2ax_{R'} |}}{a} \sinh (at_{R'})  
\end{split}
\end{equation}
\noindent for $x_{R'} \leq -\frac{1}{2a}$. Note that imposing the above conditions on the coordinate $x_{R'}$ fixes the signature of the metric, since for $x_{R'} \leq -\frac{1}{2a}$ or $1+ 2ax_{R'} \leq 0$ the metric signature changes to $(+,-)$, while for $1+ 2ax_{R'}\geq 0$ the metric has signature $(-,+)$. Thus one sees that the crossing of the horizon is achieved by the crossing of the coordinate singularity, which is precisely the tunneling barrier that causes the radiation in this formalism.
As a final comment, we note that the determinant of the metric for \eqref{rindler} is zero at the horizon $x_R = -1/a$, while
the determinant of the metric given by \eqref{rindler2} is $1$ everywhere.

\section{The WKB/Tunneling method} 

In this section we study a scalar field placed in a background metric. Physically, these fields come from the quantum fields, i.e., vacuum fluctuations, that permeate the spacetime given by the metric. In addition, as shown in reference \onlinecite{grish}, the vacuum field fluctuations obey the principle of equivalence, which supports this approach. By applying the WKB method to this scalar field, we find that the phase of the scalar field develops imaginary
contributions upon crossing the horizon. The exponential of these imaginary contributions is interpreted as a
tunneling amplitude through the horizon. By assuming a Boltzmann distribution and associating 
it with the tunneling amplitude, we obtain the temperature of the radiation.  

Writing the scalar field in terms of a phase factor as $\phi= \phi_0e^{\frac{i}{\hbar}S(t,\vec{x})}$, the Hamilton--Jacobi equations for the action $S$ of the field $\phi$ in 
the gravitational background given by the metric $g_{\mu \nu}$ are (see Appendix I for details)
\begin{equation}
\label{HJ}
g^{\mu\nu}\partial_\nu(S)\partial_\mu(S) + m^2 = 0 ~.
\end{equation}

Now for stationary spacetimes (technically spacetimes for which one can 
define a time--like Killing vector that yields a conserved energy, $E$) the action $S$ can be split into a time 
and space part, i.e., $S(t,\vec{x})= Et + S_0(\vec{x})$. 

If $S_0$ has an imaginary part, this then gives
the tunneling rate, $\Gamma_{QM}$, via the standard WKB formula. The WKB approximation tells us how to find the transmission probability in terms of the incident wave and transmitted wave amplitudes. The transition probability is in turn given by the exponentially decaying part of the wave function over the non--classical (\textit{tunneling}) region \cite{griffiths}
\begin{equation}
\label{QMrate}
\Gamma_{QM} \propto e^{-{\rm Im} \frac{1}{\hbar}\oint \; p_x dx}~.
\end{equation}
\noindent The tunneling rate given by \eqref{QMrate} is just the lowest
order, quasiclassical approximation to the full non--perturbative
Schwinger \cite{schwinger} rate. \footnote{The Schwinger rate is found by taking the Trace--Log
of the operator $(\Box _g - m^2c^2/\hbar^2)$, where $\Box _g$ is the d'Alembertian 
in the background metric $g_{\mu \nu}$, i.e., the first term in \eqref{KG}.
As a side comment, the Schwinger rate was initially calculated for the case of a uniform electric field. In this case, the Schwinger rate corresponded to the probability of creating particle--antiparticle pairs from the vacuum field at the expense of the electric field's energy. This electric field must have a critical strength in order for the Schwinger effect to occur. A good discussion of the calculation of the Schwinger rate for the usual case of a uniform electric field and the connection of the Schwinger effect to Unruh and Hawking radiation can be found in reference \onlinecite{holstein}.} 

In most cases (with an important exception that we will discuss in appendix II),
$p ^{out}$ and $p^{in}$ have the same magnitude but opposite signs. Thus $\Gamma_{QM}$ 
will receive equal contributions from the ingoing and outgoing particles, since the sign
difference between $p ^{out}$ and $p^{in}$ will be compensated for by the minus sign
that is picked up in the $p^{in}$ integration due to the fact that the path is
being traversed in the backward $x$-direction. In all quantum mechanical tunneling
problems that we are aware of this is the case: the tunneling rate across
a barrier is the same for particles going right to left or left to right. For 
this reason, the tunneling rate \eqref{QMrate} is usually written as \cite{griffiths}
\begin{equation}
\label{quasirate}
\Gamma_{QM} \propto e^{ \mp 2{\rm Im} \frac{1}{\hbar}\int \; p_x ^{out, in} dx}~,
\end{equation}
In \eqref{quasirate} the $-$ sign goes with $p_x ^{out}$ and the $+$ sign with $p_x ^{in}$.

There is a technical reason to prefer \eqref{QMrate} over \eqref{quasirate}. As was
remarked in references \onlinecite{chowdhury, akhmedov, pilling}, equation \eqref{QMrate} 
is invariant under canonical transformations, whereas the form given by \eqref{quasirate} is not. 
Thus the form given by \eqref{quasirate} is not a proper observable.
Moreover, in appendix II, we will show an example of the WKB method for
the Schwarzschild spacetime in Painlev{\'e}-Gulstrand coordinates,
and we will find that the two formulas, \eqref{QMrate} and \eqref{quasirate}, {\it are not} numerically equivalent.

However, for the case of the gravitational WKB problem, equation \eqref{quasirate} only gives the imaginary contribution to the total action coming from the spatial part of the action. In addition, there is a temporal piece, $E\Delta t$, that must be added to the total imaginary part of the action to obtain the tunneling rate. This temporal piece originates from an imaginary change of the time coordinate as the horizon is crossed. We will explicitly show how to account for this temporal piece in the next section, where we apply the WKB method to the Rindler spacetime. This imaginary part of the total action coming from the time piece is a unique feature of the gravitational WKB problem. Therefore, for the case of the gravitational WKB problem, the tunneling rate is given by
\begin{equation}
\label{gravWKBrate}
\Gamma \propto e^{-\frac{1}{\hbar}[{\rm Im} \left(\oint \; p_x dx\right)-E{\rm Im}(\Delta t)]}~.
\end{equation}
In order to obtain the temperature of the radiation, we assume a Boltzmann distribution for
the emitted particles
\begin{equation}
\label{boltz}
\Gamma \propto e^{-\frac{E}{T}}~, 
\end{equation}
where $E$ is the energy of the emitted particle,
$T$ is the temperature associated with the radiation, and we have set Boltzmann's constant, $k_B$, equal to 1. Eq. \eqref{boltz} gives the probability that a system at temperature $T$ occupies a quantum state with energy $E$. One weak point of this derivation is that we had to assume a Boltzmann distribution for the radiation while the original derivations \cite{hawking, unruh} obtain the thermal spectrum without any assumptions. Recently, this shortcoming of the tunneling method has been addressed in reference \onlinecite{banerjee-thermal}, where the thermal spectrum was obtained within the tunneling method using density matrix techniques of quantum mechanics.

By equating \eqref{boltz} and \eqref{gravWKBrate}, we obtain the following formula for the temperature $T$
\begin{equation}
\label{temperature}
T=\frac{E\hbar}{{\rm Im} \left(\oint \; p_x dx\right)-E{\rm Im}(\Delta t)} ~.
\end{equation}

\section{Unruh radiation via WKB/tunneling}

We now apply the above method to the alternative Rindler metric previously introduced.
For the $1+1$ Rindler spacetimes, the Hamilton--Jacobi equations (H--J) 
reduce to $g^{tt}\partial_tS\partial_tS + g^{xx}\partial_xS\partial_xS + m^2 =0$.
For the Schwarzschild--like form of Rindler given in \eqref{rindler2} the H--J equations are
\begin{equation}
\label{HJR2}
-\frac{1}{(1 + 2\,a\,x_{R'})}(\partial_t S)^2 + (1 + 2\,a\,x_{R'})(\partial_x S)^2 +m^2 =0~.
\end{equation} 

Now splitting up the action $S$ as $S(t,\vec{x})= Et + S_0(\vec{x})$ in \eqref{HJR2}
gives
\begin{equation}
\label{HJR2a}
-\frac{E}{(1 + 2\,a\,x_{R'})^2} + (\partial_x S_0 (x_{R'}))^2 +
\frac{m^2}{1 + 2\,a\,x_{R'}} =0~.
\end{equation} 
From \eqref{HJR2a}, $S_0$ is found to be
\begin{equation}
\label{HJR2b}
S^{\pm}_0 =\pm \int_{-\infty}^\infty \frac{\sqrt{E^2 - m^2(1+2\,a\,x_{R'})}}{1+2\,a\,x_{R'}} ~dx_{R'}~.
\end{equation}
In \eqref{HJR2b}, the $+$ sign corresponds to the ingoing particles
(i.e., particles that move from right to left) and the $-$ sign to the outgoing particles (i.e., particles
that move left to right). Note also that \eqref{HJR2b} is of the 
form $S_0 = \int p_x  ~dx$, where $p_x$ is the canonical momentum of the field in 
the Rindler background. The Minkowski spacetime expression for the momentum is easily 
recovered by setting $a=0$, in which case one sees that $p_x = \sqrt{E^2 -m^2}$.

From \eqref{HJR2b}, one can see that this integral has a pole along the path of integration at $x_{R'} = -\frac{1}{2a}$.
Using a contour integration gives an imaginary contribution to the action. We will give explicit details
of the contour integration since this will be important when we try to apply this method to the standard
form of the Rindler metric \eqref{rindler} (see Appendix III for the details of this calculation). We go around the pole at $x_{R'}=-\frac{1}{2a}$ using a semi--circular
contour which we parameterize as $x_{R'}= -\frac{1}{2a} + \epsilon e ^{i \theta}$, where $\epsilon \ll 1$ and $\theta$ 
goes from $0$ to $\pi$ for the ingoing path and $\pi$ to $0$ for the outgoing path. These contours are illustrated in the figure below. With this parameterization
of the path, and taking the limit $\epsilon \rightarrow 0$, we find that the imaginary part of \eqref{HJR2b}
for ingoing ($+$) particles is
\begin{equation}
\label{rate2}
S^{+}_0 =  \int^{\pi}_{0} \frac{\sqrt{E^2 - m^2 \epsilon e^{i \theta}}}{2 a \epsilon e^{i \theta}} ~
i \epsilon e^{i \theta} d \theta =  \frac{i \, \pi \, E}{2a} ~,
\end{equation}
and for outgoing ($-$) particles, we get

\begin{equation}
S^{-}_0 = -\int^{0}_{\pi} \frac{\sqrt{E^2 - m^2 \epsilon e^{i \theta}}}{2 a \epsilon e^{i \theta}} ~
i \epsilon e^{i \theta} d \theta = \frac{i \, \pi \, E}{2a} ~.
\end{equation}

\begin{figure} [H]
  \centering\includegraphics[trim = 40mm 169mm 65mm 78mm, clip, width=8.5cm]{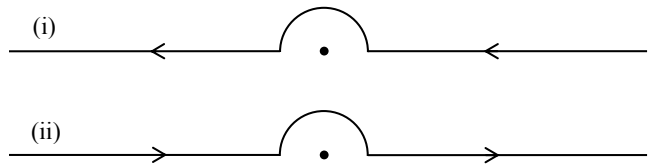}
\noindent\caption{Contours of integration for (i) the ingoing and (ii) the outgoing particles.}
\label{fig:contour}
\end{figure}

%Using the results from \eqref{rate2} in \eqref{temperature} (and taking into account the
%additional minus sign coming from the backward traversal of the path for the ingoing particle
%in \eqref{QMrate}), one obtains a temperature for Unruh radiation
%of $T= (E\hbar)/\left[\frac{\pi \, E}{2a} + \frac{\pi \, E}{2a} \right]= \frac{\hbar a}{\pi}$.
%This is twice of the value of the Unruh temperature, $T_{Unruh} = \frac{\hbar a}{2 \pi}$.

In order to recover the Unruh temperature, we need to take into account the contribution from the time piece of the total action $S(t,\vec{x})= Et + S_0(\vec{x})$, as indicated by the formula of the temperature, \eqref{temperature}, found in the previous section.
The transformation of \eqref{coordtrans1} into \eqref{coordtrans2} indicates that the time coordinate has a discrete
imaginary jump as one crosses the horizon at $x_{R'} = -\frac{1}{2a}$, since the two time coordinate
transformations are connected across the horizon by the change
$t_{R'} \rightarrow t_{R'} - \frac{i\pi}{2a}$, that is, 
$$\sinh (at_{R'}) \rightarrow \sinh\left( at_{R'}-\frac{i\pi}{2} \right)= - i\cosh(at_{R'})~.$$ Note that as the horizon is crossed, a factor of $i$ comes from the term in front of the hyperbolic function in \eqref{coordtrans1}, i.e.,
$$ \sqrt{1+2ax_{R'}}\rightarrow i\sqrt{|1+2ax_{R'}|}~,$$ so that \eqref{coordtrans2} is recovered. 

Therefore every time the horizon is crossed,
the total action $S(t,\vec{x})=S_0(\vec{x})+Et$ picks up a factor of $E \Delta t = -\frac{i\pi E}{2a}$. For the
temporal contribution, the direction in which the horizon is crossed 
does not affect the sign. This is different from the situation for the spatial contribution. 
When the horizon is crossed once, the total action $S(t,\vec{x})$
gets a contribution of $E \Delta t = -\frac{i E\pi}{2a}$, and for a round trip,
as implied by the spatial part $\oint \; p_x dx$, the total contribution is $E \Delta t_{total} = -\frac{i E\pi}{a}$.
So using the equation for the temperature \eqref{temperature} developed in the previous section, we obtain
\begin{equation}
\label{Unruh-temp}
T_{Unruh} = \frac{E\hbar}{\frac{\pi E}{a} + \frac{\pi E}{a}} = \frac{\hbar a}{2\pi}~,
\end{equation}
which is the Unruh temperature. The interesting feature of this result is
that the gravitational WKB problem has contributions from both spatial and
temporal parts of the wave function, whereas the ordinary quantum mechanical WKB
problem has only a spatial contribution. This is natural since
time in quantum mechanics is treated as a distinct parameter, separate in character from
the spatial coordinates. However, in relativity time is on equal footing with the
spatial coordinates.

\section{Conclusion} 

We have given a derivation of Unruh radiation in terms of the original heuristic explanation 
as tunneling of virtual particles tunneling through the horizon \cite{hawking}. 
This tunneling method can easily be applied to different spacetimes and
to different types of virtual particles. We chose
the Rindler metric and Unruh radiation since, because of the
local equivalence of acceleration and gravitational fields, it represents the prototype
of all similar effects (e.g. Hawking radiation, Hawking--Gibbons radiation). 

Since this derivation touches on many different areas -- classical mechanics (through the H--J equations), relativity (via the Rindler metric), relativistic field theory (through the Klein--Gordon equation in curved backgrounds),
quantum mechanics (via the WKB method for gravitational fields), thermodynamics (via the Boltzmann distribution
to extract the temperature), and mathematical methods (via the contour integration to obtain
the imaginary part of the action) -- this single problem serves as a reminder of the
connections between the different areas of physics. 

This derivation also highlights several subtle points regarding the Rindler metric and the WKB
tunneling method. In terms of the Rindler metric, we found that the different forms of the
metric \eqref{rindler} and \eqref{rindler2} do not cover the same parts of the full spacetime
diagram. Also, as one crosses the horizon, there is an imaginary jump of the Rindler time coordinate
as given by comparing \eqref{coordtrans1} and \eqref{coordtrans2}. 

In addition, for the gravitational WKB problem, $\Gamma$ has contributions from both the spatial and 
temporal parts of the action. Both these features are not found in the ordinary quantum
mechanical WKB problem. 

As a final comment, note that one can define an absorption probability (i.e., $P_{abs} \propto |\phi _{in}|^2$)
and an emission probability (i.e., $P_{emit} \propto |\phi _{out}|^2$). These probabilities can
also be used to obtain the temperature of the radiation via the ``detailed balance method" \cite{padman} 
$$\frac{P_{emit}}{P_{abs}} = e^{-E/T}~.$$ 
Using the expression of the field $\phi= \phi_0e^{\frac{i}{\hbar}S(t,\vec{x})}$, the Schwarzschild--like form of the Rindler metric given in \eqref{rindler2}, 
and taking into account the spatial {\it and} temporal contributions gives an an absorption probability of
$$P_{abs} \propto e^{\frac{\pi E}{a} -\frac{\pi E}{a}}=1$$ 
and an emission probability of
$$P_{emit} \propto e^{-\frac{\pi E}{a} -\frac{\pi E}{a}} = e^{-\frac{2 \pi E}{a}}~.$$ 
The first term in the exponents of the above probabilities corresponds to the 
spatial contribution of the action $S$, while the second term is the time
piece. When using this method, we are not dealing with a directed line integral as in \eqref{QMrate}, so the spatial
parts of the absorption and emission probability have opposite signs. In addition, the absorption 
probability is $1$, which physically makes sense -- particles should be able to fall into the horizon 
with unit probability. If the time part were not included in $P_{abs}$, then for some given $E$ and $a$
one would have $P_{abs} \propto e^{\frac{\pi E}{a}}>1$, i.e., the probability of absorption would exceed $1$
for some energy. Thus for the detailed balance method the temporal piece is crucial to ensure that one has a
physically reasonable absorption probability.

\begin{center}
{\bf Acknowledgments}
\end{center}
The authors would like to thank E.T. Akhmedov for valuable discussions. D.S. is supported by a
2008-2009 Fulbright Scholars Grant. D.S. would like to thank Vitaly Melnikov for
the invitation to research at the Center for Gravitation and Fundamental
Metrology and the Institute of Gravitation and Cosmology at PFUR. The authors would also like to thank two anonymous referees from this journal for their valuable comments and suggestions that led to the final version of this paper.

\begin{center}
{\bf Appendix I: The Hamilton--Jacobi equations}
\end{center}
The Hamilton--Jacobi equations may be derived from the Klein--Gordon equation in the following manner.
The Klein--Gordon (KG) equation for a scalar field $\phi$ of mass $m$, placed in a
background metric $g_{\mu \nu}$ is
\begin{equation}
\label{KG}
\left(\frac{1}{\sqrt{-g}}\partial_\mu(\sqrt{-g}g^{\mu\nu}\partial_\nu)- \frac{m^2c^2}{\hbar^2}\right)\phi=0~,
\end{equation}
where $c$ is the speed of light and $\hbar$ is Planck's constant.
For Minkowski spacetime, the above reduces to the free Klein--Gordon equation, i.e.,
 $(\Box - m^2c^2/\hbar^2)\phi=(-\partial^2/c^2\partial t^2 + \nabla^2 - m^2c^2/\hbar^2)\phi=0$. 

In using a scalar field, we are 
following the original works \cite{hawking, unruh}. Despite the fact that,
absent the hypothetical Higgs boson, there are no known fundamental scalar fields,
the derivation with spinor or vector particles would only add the 
complication of having to carry around spinor or Lorentz indices without adding
to the basic understanding of the phenomenon. Using the WKB approach presented here it
is straightforward to do the calculation using spinor\cite{mann} or vector particles. 
 
Setting the speed of light $c$ equal to $1$, multiplying \eqref{KG} by $-\hbar$ and using the product rule, \eqref{KG} becomes
\begin{equation}
\label{KGb}
\begin{split}
&\frac{-\hbar^2}{\sqrt{-g}}\Big[ (\partial_\mu\sqrt{-g}) g^{\mu\nu} \partial_\nu\phi
+ \sqrt{-g}(\partial_\mu g^{\mu\nu})\partial_\nu\phi + \\ 
&\sqrt{-g}g^{\mu\nu}\partial_\mu\partial_\nu\phi \Big] + m^2\phi = 0 ~.
\end{split}
\end{equation}

\noindent The above equation can be simplified using the fact that the covariant 
derivative of any metric $g$ vanishes  
\begin{equation}
\label{vanishcov}
\nabla_{\alpha}g^{\mu\nu} = \partial_\alpha g^{\mu\nu}+ \Gamma^\mu_{\alpha\beta}g^{\beta\nu}+ \Gamma^\nu_{\alpha\beta}g^{\mu\beta} = 0~,
\end{equation}
where $\Gamma^\mu_{\alpha\beta}$ is the Christoffel connection.
All the metrics that we consider here are diagonal so $\Gamma^\mu_{\alpha\beta} = 0$, for $\mu\not=\alpha\not=\beta$. 
It can also be shown that
\begin{equation}
\label{gamma}
\Gamma^\mu_{\mu\gamma} = \partial_\gamma (\ln \sqrt{-g})= \frac{\partial_\gamma \sqrt{-g}}{\sqrt{-g}} ~.
\end{equation}
Using \eqref{vanishcov} and \eqref{gamma}, the term $\partial_\mu g^{\mu\nu}$ in 
\eqref{KGb} can be rewritten as
\begin{equation}
\label{dgterm}
\partial_\mu g^{\mu\nu} =  -\Gamma^\mu_{\mu\gamma}g^{\gamma\nu} - \Gamma^\nu_{\mu\rho}g^{\mu\rho}
= -\frac{\partial_\gamma \sqrt{-g}}{\sqrt{-g}} g^{\gamma\nu} ~,										
\end{equation}
since the harmonic condition is imposed on the metric $g^{\mu\nu}$, i.e., $\Gamma^\nu_{\mu\rho}g^{\mu\rho}=0$. Thus \eqref{KGb} becomes
\begin{equation}
\label{KGc}
-\hbar^2g^{\mu\nu}\partial_\mu\partial_\nu\phi + m^2\phi = 0 ~.
\end{equation}
We now express the scalar field $\phi$ in
terms of its action $S=S(t,\vec{x})$ 
\begin{equation}
\phi= \phi_0e^{\frac{i}{\hbar}S(t,\vec{x})} ~,
\end{equation}
\noindent where $\phi_0$ is an amplitude \cite{landau} not
relevant for calculating the tunneling rate. Plugging this expression for $\phi$ into \eqref{KGc}, we get
\begin{equation}
-\hbar g^{\mu\nu}(\partial_\mu(\partial_\nu(iS))) + g^{\mu\nu}\partial_\nu(S)\partial_\mu(S) + m^2 = 0 ~.
\end{equation}
\noindent Taking the classical limit, i.e., letting $\hbar\rightarrow 0$, 
we obtain the Hamilton--Jacobi equations for the action $S$ of the field $\phi$ in the gravitational background given by the metric $g_{\mu \nu}$,
\begin{equation}
g^{\mu\nu}\partial_\nu(S)\partial_\mu(S) + m^2 = 0 ~.
\end{equation}

\begin{center}
{\bf Appendix II: Hawking radiation from the Painlev{\'e}--Gulstrand form of the Schwarzschild spacetime}
\end{center}
The Painlev{\'e}--Gulstrand form of the Schwarzschild spacetime is obtained by transforming the Schwarzschild time
$t$ to the  Painlev{\'e}--Gulstrand time $t'$
\begin{equation} 
\label{pg-trans}
dt = dt' - \frac{\sqrt{\frac{2M}{r}} \, dr}{1-\frac{2\,M}{r}}.
\end{equation} 
Applying the above transformation to the Schwarzshild metric gives the Painlev{\'e}--Gulstrand 
form of the Schwarzschild spacetime
\begin{equation} 
\label{pmetric}
ds^2 = - \left(1-\frac{2M}{r}\right)\, {dt'}^2 +
2\sqrt{\frac{2M}{r}}\, dr\,dt' + dr^2 ~.
\end{equation}
The time is transformed, but all the other coordinates ($r, \theta , \phi$) are
the same as the Schwarzschild coordinates.
If we use the metric in \eqref{pmetric} to calculate the spatial part of the action as in \eqref{HJR1b}
and \eqref{HJR2b}, we obtain
\begin{eqnarray} 
\label{ft}
S_0 &=& - \int_{-\infty} ^{\infty}
\frac{dr}{1-\frac{2M}{r}}\,\sqrt{\frac{2M}{r}}\, E \\
&\pm& \int_{-\infty} ^{\infty}
\frac{dr}{1-\frac{2M}{r}}\,\sqrt{E^2 - m^2 \left(1 -
\frac{2M}{r}\right)}.
\end{eqnarray} 
Each of these two integrals has an imaginary contribution of equal magnitude, as can be seen
by performing a contour integration.
Thus one finds that for the ingoing particle (the $+$ sign in the second integral) one has a zero
net imaginary contribution, while from for the outgoing particle (the $-$ sign in the second integral)
there is a non--zero net imaginary contribution. Therefore in this case there is clearly a difference
by a factor of two between using \eqref{QMrate} and \eqref{quasirate} which comes exactly because the
tunneling rates from the spatial contributions in this case do depend upon the direction in 
which the barrier (i.e., the horizon) is crossed. The Schwarzcshild metric has a similar temporal contribution
as for the Rindler metric \cite{akhmedov2}. The Painlev{\'e}--Gulstrand form of the Schwarzschild metric actually has {\em two} temporal contributions:
(i) one coming from the jump in the Schwarzschild time coordinate similar to what occurs with the Rindler
metric in \eqref{coordtrans1} and \eqref{coordtrans2}; (ii) the second temporal contribution coming from the transformation between the Schwarzschild and Painlev{\'e}--Gulstrand time coordinates in \eqref{pg-trans}. If one integrates
equation \eqref{pg-trans}, one can see that there is a pole coming from the second term. One needs
to take into account both of these time contributions in addition to the spatial contribution, 
to recover the Hawking temperature. Only by adding the temporal contribution to the spatial part from \eqref{QMrate}, does one recover the Hawking temperature \cite{akhmedov2}
$T=\frac{\hbar}{8 \pi M}$. Thus for both reasons -- canonical invariance and to recover the 
temperature -- it is \eqref{QMrate} which should be used over \eqref{quasirate}, 
when calculating $\Gamma_{QM}$. In ordinary quantum mechanics, there is never a case -- as far as we know -- where it
makes a difference whether one uses \eqref{QMrate} or \eqref{quasirate}. This feature -- dependence of the
tunneling rate on the direction in which the barrier is traverse -- appears to be a unique feature of the
gravitational WKB problem. So in terms of the WKB method as applied to
the gravitational field, we have found that there are situations (e.g. Schwarzschild spacetime
in  Painlev{\'e}--Gulstrand coordinates) where the tunneling rate depends on 
the direction in which barrier is traversed so that \eqref{QMrate} over \eqref{quasirate} are not equivalent
and will thus yield different tunneling rates, $\Gamma$. 

\begin{center}
{\bf Appendix III: Unruh radiation from the standard Rindler metric}
\end{center}
For the standard form of the Rindler metric given by \eqref{rindler},
the Hamilton--Jacobi equations become
\begin{equation}
\label{HJR1}
-\frac{1}{(1+a\,x_R)^2}\, ( \partial_t S )^2 + (\partial_x S )^2 + m^2 =0~.
\end{equation}
After splitting up the action as $S(t,\vec{x})= Et + S_0(\vec{x})$, we get
\begin{equation}
\label{HJR1a}
-\frac{E}{(1+a\,x_R)^2}\, + (\partial_x S_0 (x_R) )^2 + m^2 =0~.
\end{equation}
The above yields the following solution for $S_0$
\begin{equation}
\label{HJR1b}
S^{\pm}_0 = \pm \int_{-\infty}^\infty \frac{\sqrt{E^2 -
m^2\, (1 + a\, x_R)^2}}{1 + a\,x_R}\, dx_R~,
\end{equation}
where the $+ (-)$ sign corresponds to the ingoing (outgoing) particles.

Looking at \eqref{HJR1b}, we see that the pole is now at $x_R = -1/a$ and a naive application of contour integration appears to give the results $\pm \frac{i \, \pi \, E}{a}$. However, this cannot be justified since
the two forms of the Rindler metric -- \eqref{rindler} and \eqref{rindler2} -- are related by 
the simple coordinate transformation \eqref{coordtrans}, and one should not change
the value of an integral by a change of variables. The resolution to this puzzle is that
one needs to transform not only the integrand but the path of integration, so applying
the transformation \eqref{coordtrans} to the semi--circular contour
$x_{R'}= -\frac{1}{2a} + \epsilon e ^{i \theta}$ gives $x_R = -\frac{1}{a}+\frac{\sqrt{\epsilon}}{a} e^{i \theta /2}$.
Because $e ^{i \theta}$ is replaced by $e^{i \theta /2}$ due to the square root in the 
transformation \eqref{coordtrans}, the semi--circular contour of \eqref{rate2} is replaced by
a quarter--circle, which then leads to a contour integral of $i \frac{\pi}{2} \times {\rm Residue}$ instead of
$i \pi \times {\rm Residue}$. Thus both forms of Rindler yield the same spatial contribution to the total imaginary part of the action.
 
%%%%%%%%%%%%%%%%%%%%%%%%%%%%%%%%%%%%%%%%%%% 

\end{document}